\documentclass[12pt]{article}

\begin{document}
\setcounter{page}{0}
\def\hatS{\hat s}
\def\hatT{\hat t}
\def\hatU{\hat u}
\def\cW{\cos \theta_W}
\def\sW{\sin \theta_W}
\def\warp{{\cal K} R_c \phi}
\def\warpp{{\pi {\cal K} R_c}}

%========================================================================%
\thispagestyle{empty}
\begin{flushright}
                                                   IITK-HEP-00-01  \\
                                                   hep-ph/0002079 \\
\end{flushright} 
\vskip 20pt
\begin{center}
{\large\bf  
HERA Constraint on Warped Quantum Gravity}
\vskip 20pt
{\sl  
Prasanta Das~\footnote{E-mail: pdas@iitk.ac.in}, 
Sreerup Raychaudhuri~\footnote{E-mail: sreerup@iitk.ac.in} {\rm and}
Saswati Sarkar }
\vskip 10pt
{\rm
Department of Physics, Indian Institute of Technology, \\
Kanpur 208 016, India.} \\
\vskip 25pt
{\large\bf Abstract} 
\end{center} 
\vskip 5pt
\noindent
We study recent data on deep inelastic $e^+ p$ scattering at HERA
to constrain the parameters of a Randall-Sundrum-type scenario of 
quantum gravity with a small extra dimension and a non-factorable 
geometry.  
\vskip 100pt 
\begin{flushleft}
                 February 2000
\end{flushleft} 

%========================================================================%
\newpage 

Theories with extra dimensions which predict observable consequences
at the current high energy accelerators have lately attracted a great
deal of interest. Following the original suggestion by Arkani-Hamed,
Dimopoulos and Dvali (ADD)~\cite{ArHa-Dimo-Dval}, 
there have been numerous studies in the
literature~\cite{Rizz-Revw} which probe consequences of 
multiple Kaluza-Klein graviton exchange
leading to interactions of electroweak strength. The fact that these
theories predict quantum gravity effects at TeV scales has been 
suggested~\cite{ArHa-Dimo-Dval}
as a solution of the well-known hierarchy problem in the Standard Model 
(SM). Though novel and interesting, however, the model suggested by ADD,
which is based on a factorable ${\bf R}^4 \times ({\bf S}^1)^d$ geometry, 
$d$ being
the number of extra compact dimensions, has the drawback of introducing 
large compactification radii (amounting to an energy scale as low as 
$10^{-13}$ GeV), which effectively introduces a new hierarchy
problem. Motivated by this, Randall and Sundrum (RS)~\cite{Rand-Sund} 
have suggested a 
somewhat different mechanism to solve the hierarchy problem. Instead
of writing a factorable metric
\begin{equation}
ds^2 = \eta_{\mu\nu} ~dx^{\mu} ~dx^{\nu} ~+~ R_c^2 ~d\phi_i ~d\phi_i
 \label{metric_ADD}
\end{equation}
where the $\phi_i~(i = 1,d)$ are extra dimensions compactified with a
common radius $R \sim 1$ mm, they write a {\em non-factorable} metric
\begin{equation}
ds^2 = e^{-\warp} ~\eta_{\mu\nu} ~dx^{\mu} dx^{\nu} ~+~ R_c^2 ~d\phi^2
 \label{metric_RS}
\end{equation}
involving one extra dimension compactified with a radius $R_c$,
which is assumed to be marginally greater than the Planck length
$10^{-33}$ cm, and an extra mass scale ${\cal K}$, which is related to 
the Planck scale $M^{(5)}_P$ in the five-dimensional bulk by 
${\cal K} \left[M^{(4)}_P\right]^2 \simeq \left[M^{(5)}_P\right]^3$. 
Such a `warped' geometry
is motivated by compactifying the extra dimension on a 
${\bf S}^1/{\bf Z}_2$ orbifold,
with two $D$-branes at the orbifold fixed points, {\em viz.}, one at 
$\phi = 0$ (`Planck brane' or `invisible brane'),
and one at $\phi = \pi$ (`TeV brane' or `visible brane'). 
It can then be shown that if we assume matter fields 
to be confined to these $D$-branes, one can solve the 
Einstein equations to obtain a metric of the above
form. The interesting physical consequence of this geometry is
that any mass scale ${\cal M}$ on either brane gets scaled by the `warp
factor $e^{-\warp}$ on either brane. Thus, a mass scale on the Planck
brane ($\phi = 0$) will remain unchanged, but any mass scale on the
TeV brane will be scaled by a factor $e^{-\warpp}$. If we assume that 
the Planck scale is the only fundamental mass scale in the theory, all
masses on the TeV brane will be scaled to
\begin{equation}
{\cal M} \sim e^{-\warpp} M^{(4)}_P
  \label{planckmass}
\end{equation}
It now requires ${\cal K} R_c \simeq 11 - 12$ to obtain ${\cal M}$ of the 
order of the electroweak scale, which justifies the name `TeV brane'. 
Thus, in this model there is no 
hierarchy problem, since all the independent mass scales are close
to the Planck scale. There still remains a minor problem: that of
stabilizing the radius $R_c$ (which is marginally smaller than the
Planck scale) against quantum fluctuations, but this
is not so severe as in the model of ADD, where the compactification
radius needs to be stabilized over as many as 30 orders of magnitude. 
A simple extension of the RS construction involving an extra bulk scalar
field has been proposed~\cite{Gold-Wise-2} to stabilize $R_c$
and this predicts light
radion excitations with possible collider signatures~\cite{Maha-Raks}.
However, as these will not contribute to the processes of interest 
in this
letter, this idea will not be discussed further.  On the flip side, 
it is not as simple to embed the RS construction within the
framework of string theories as it is for the ADD case. However,
a first attempt has been made~\cite{Verl}, and it may be hoped that 
future work will achieve this highly desirable goal. 

Following the ingenious suggestion of a non-factorable geometry, 
the mass spectrum and couplings of the graviton in the RS model have 
been worked out,
in Refs.~\cite{Gold-Wise,Davo-Hewt-Rizz}. We do not describe the details
of this calculation, but refer the reader to the original literature.
It is worth noting that there are strong phenomenological constraints
on bulk excitations of the SM fields~\cite{Davo-Hewt-Rizz-2}.
It suffices here to note that the effective Lagrangian density for 
graviton interactions on the TeV brane (which we identify with the
observable world) has the form~\cite{Davo-Hewt-Rizz}
\begin{equation}
{\cal L}_{eff}^{RS} = 
- \frac{1}{\overline{M}_P} ~h^0_{\mu\nu}(x) ~T^{\mu\nu}(x)
- \frac{e^{\warpp}}{\overline{M}_P} \sum_{n=1}^\infty 
~h^n_{\mu\nu}(x) ~T^{\mu\nu}(x)
 \label{lagrangian}
\end{equation}
where $\overline{M}_P \equiv M^{(4)}_P/ \sqrt{8\pi}$ is the reduced Planck 
mass and the $h^n_{\mu\nu}(x)$ correspond to the Kaluza-Klein expansion 
of the massless graviton in five dimensions
\begin{equation} 
h_{\mu\nu}(x, \phi) ~=~ 
\sum_{n=0}^\infty ~h^n_{\mu\nu}(x) ~\frac{\chi^n(\phi)}{\sqrt{R_c}} \ .
 \label{gravitonexpansion}
\end{equation} 
Equation (\ref{lagrangian}) tells us that the massless Kaluza-Klein (KK)
mode effectively decouples from ordinary matter since its interactions are 
suppressed by the Planck mass. On the other hand, the massive KK modes
couple as the inverse of the Planck mass, scaled by $e^{-\warpp}$, which
is an electroweak-strength interaction. Feynman rules to the lowest 
order for these
modes, assuming a coupling $\overline{M}_P^{-1}$ to {\em all} the modes
have been worked out in Refs.~\cite{Han-Lykk-Zhng})
and~\cite{Giud-Ratt-Well} in the context of ADD-like scenarios. 
All we need to do to get the corresponding Feynman rules in the RS model
is to multiply the couplings by the warp factor $e^\warpp$ where 
necessary. 

As shown in Ref.~\cite{Gold-Wise}, the orbifold geometry forces the Fourier 
coefficients $\chi^n(\phi)$ to satisfy a Bessel equation, whence it may be 
shown that they are given by a linear combination 
of the Bessel and Neumann functions of order 2. 
The requirement that the
first derivative of $\chi^n(\phi)$ be continuous at the orbifold fixed point 
$\phi = \pi$ then requires $J_1(x_n)  = 0$. Using this, 
the masses $M_n$ of the graviton states can be written in terms of the zeros
of the Bessel function of order unity as 
\begin{equation}
M_n = x_n {\cal K} e^{-\warpp}  \equiv  x_n m_o
   \label{mass}
\end{equation}
where $m_0$ sets the scale of graviton masses and is 
essentially a free parameter of the theory. It is also
convenient to write 
\begin{equation}
\frac{e^{\warpp}}{\overline{M}_P} = \frac{c_0}{m_0} ~\sqrt{8\pi} 
  \label{coupling}
\end{equation}
using (\ref{mass}) and 
introducing another undetermined parameter $c_0 \equiv {\cal K}/M^{(4)}_P$. 
Ref.~\cite{Davo-Hewt-Rizz} points out that ($m_0, c_0$)  may conveniently be
taken as the free parameters of the theory, and we follow their 
prescription in our work. 

Though $c_0$ and $m_0$ are not precisely known,
one can make estimates of their magnitude using theoretical ideas and
phenomenological inputs. We note that the RS construction requires ${\cal K}$
to be at least an order of magnitude less than $M^{(4)}_P$,
because ${\cal K}^{-1}$ 
sets the scale for the curvature of the fifth dimension,
and should therefore be large compared with the Planck length. The
latter is necessitated by the requirement that 
fluctuations in the bulk gravitational field in the vicinity of
the $D$-branes be small. The range of
interest for $c_0$ is, therefore, about 0.01 to 0.1, the lower value being 
determined by naturalness considerations. Regarding $m_0$, Eq. (\ref{mass})
tells us that it is reduced from the scale ${\cal K}$ by the factor
$e^{-\warpp}$. In the RS construction, one requires 
${\cal K} R_c \sim 11$--12,
which reduces $m_0$ to the electroweak scale. Hence, we may consider 
$m_0$ in the range of a few tens of GeV to a few TeV. Eq.~(\ref{mass})
also tells us  
that the first massive graviton lies at $M_1 = x_1 m_0 \simeq 3.83~m_0$.
Since no graviton resonances have been seen at LEP-2, running at energies
upto 200 GeV, it is clear that we should expect $m_0 > 52$ GeV.  

In this letter we report on a study of graviton effects, within the RS model,
on $e^+ p$ deep inelastic scattering (DIS) 
at HERA. At the leading order, there are 
two extra Feynman diagrams contributing to $e^+ p \to e^+ ~+~ X$. 
One of these
involves a $t$-channel exchange of a virtual (massive) graviton between
the $e^+$ and a quark; the other involves a $t$-channel exchange of a 
virtual (massive) graviton between the $e^+$ and a gluon. The first
one adds coherently with the corresponding SM diagrams with photon
and $Z$-boson exchange; the second one has no SM analogue and hence
adds incoherently. However, at HERA energies, we do not expect much 
contribution from the gluon-induced diagram because of the low gluon flux. 

The cross-section for the above processes has been calculated for the
case of the ADD model in Ref.~\cite{Math-Rayc-Srid} and can be easily
translated to the RS model using the replacement
\begin{equation}
\frac{\lambda}{M_S^4} 
\longrightarrow \frac{8\pi c_0^2}{m_0^2} \sum_n \frac{1}{|\hatT| + M_n^2}
  \label{replacement}
\end{equation}
We have developed approximate analytic formulae for this sum, 
using the well-known
properties of the zeros of the Bessel function $J_1(x)$. These will be
presented elsewhere~\cite{PDas-Rayc-Sark}. Using this, we
incorporate the calculated theoretical cross-section into a parton-level
Monte Carlo event generator, with two free parameters, {\em viz.} the
graviton mass scale $m_0$, and the coupling parameter $c_0$. Finally 
the simulation results are compared with data from the ZEUS Collaboration 
to constrain the ($m_0$--$c_0$) plane. 

Our numerical studies are founded   
on the latest results presented~\cite{ZEUS} by the ZEUS 
Collaboration, which are based on 47.7 pb$^{-1}$ of data collected over the
period 1994-1997. The ZEUS Collaboration uses the double-angle (DA) method
to determine the DIS variables. In this method, one
measures the polar angle $\theta_e$ of the scattered positron, and 
reconstructs the polar angle $\gamma_h$ of the struck quark in the naive
parton model using all hadronic clusters
which can be identified with a jet having the requisite $p_T$ balance
with the positron.  In terms of these observables and the energy 
$E_e (E_p)$ of the initial positron (proton) beam, one can reconstruct 
the standard DIS variables as
\begin{eqnarray}
Q^2_{DA} & = & 
4 E_e^2 \frac{ \sin \gamma_h ( 1 + \cos \theta_e )}
             { \sin \gamma_h + \sin \theta_e - \sin( \gamma_h + \theta_e )}
\\
x_{DA} & = & \frac{E_e}{E_p} 
        ~\frac{ \sin \gamma_h + \sin \theta_e + \sin( \gamma_h + \theta_e )}
             { \sin \gamma_h + \sin \theta_e - \sin( \gamma_h + \theta_e )}
\\
y_{DA} & = & 
\frac{ \sin \theta_e ( 1 - \cos \gamma_h )}
             { \sin \gamma_h + \sin \theta_e - \sin( \gamma_h + \theta_e )}
\end{eqnarray}
Various triggers, acceptances and selection cuts have
been used by the ZEUS Collaboration, of which we need to impose
only the following in a parton-level analysis:
\begin{itemize}
\item If the final state positron has polar angle greater than $17.2^0$,
it must have a total energy greater than 10 GeV.
\item If the final state positron has polar angle less than $17.2^0$,
it must have a transverse momentum greater than 30 GeV. 
\end{itemize}

Of the DIS variables listed above, 
it is well-known that it is the first, namely $Q^2_{DA}$, which 
exhibits maximum sensitivity to most kinds of new physics. 
The ZEUS Collaboration has
presented their data for 20 bins in $Q^2_{DA}$, ranging from 
$Q^2_{DA}$ = 400 GeV$^2$ to 51200 GeV$^2$. The Born-level cross-section 
in each bin, obtained by suitably scaling out radiative effects, 
has been presented by the ZEUS Collaboration together with the SM 
expectation. We have checked that the
latter, obtained using hadronization procedures incorporated in the 
standard HERACLES and ARIADNE program packages, 
are in excellent agreement (within a few per cent) with our parton-level
analysis. Any small differences which persist can be removed
by calibrating the cross-section binwise, so as to yield the actual ZEUS
expectations. This procedure has the added merit of taking care of 
residual higher
order effects such as initial state radiation, which should be roughly the 
same in the SM as in the case when the graviton exchanges are 
included.  

In Fig. 1, we present a graph showing the variation in the $Q^2_{DA}$ 
distribution with the mass scale $m_0$ in the RS model.    
For this graph, we have plotted the ratio 
\begin{equation}
R (m_0,c_0) ~~\equiv~~ \frac{d\sigma_{RS}/dQ^2_{DA}}
                       {d\sigma_{SM}/dQ^2_{DA}}
  \label{ratio}
\end{equation}
of the cross-section
predicted in the RS model with the prediction of the SM, 
for $c_0$ = 0.1 and $m_0 = 80, 90$ and 100 GeV, together
with the ZEUS data. It may be seen that the cross-section in the RS model 
(like the ADD
model~\cite{Math-Rayc-Srid}) exhibits large deviations in the highest
$Q^2$ bins. This, of course, rapidly approaches the SM (dotted line)
if $c_0$ is chosen smaller. Interestingly, the RS model predictions
seem to show a slight diminution for 
intermediate values of $Q^2_{DA}$, which are intriguingly like the trend
shown by the data. However, the experimental errors are too large to
enable us to attach any significance to this circumstance.
Accordingly, we take the conservative viewpoint that the data fit 
the SM very well and can be used to constrain new physics. 

% ------------------------------------------------------------------
\begin{figure}[htb]
\begin{center}
\vspace*{2.2in}
      \relax\noindent\hskip -4.4in\relax{\includegraphics{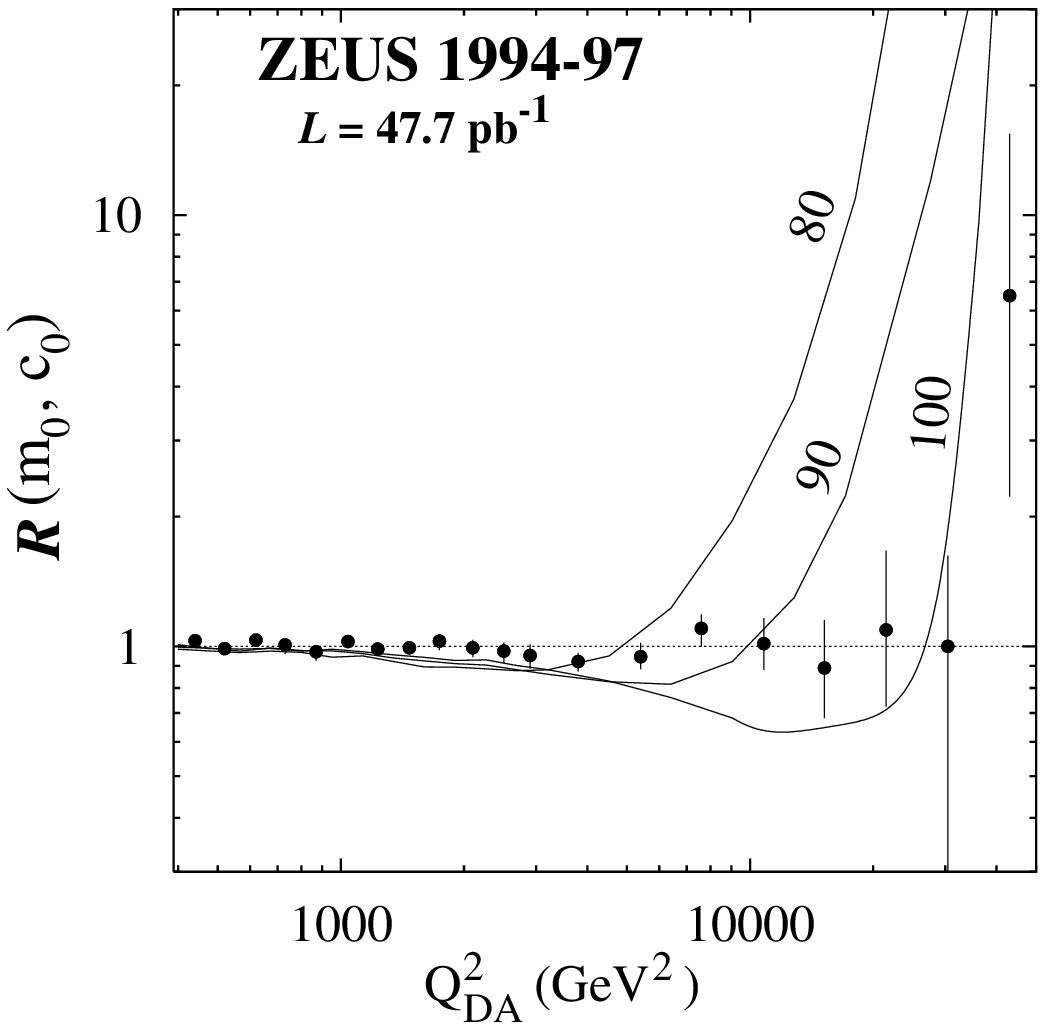}}
\end{center}
\end{figure}
\vspace*{1.5in}
\noindent {\bf Figure 1}.
{\footnotesize\it Illustrating the ratio $R(m_0,c_0)$ of the $Q^2_{DA}$ 
distribution in the RS model to that in the SM (see Eq. \ref{ratio}),
for $c_0 = 0.1$ and $m_0 = 80, 90$ and 100 GeV. The dotted line 
corresponds to the SM. The ZEUS data are also shown. }
% ------------------------------------------------------------------
\vskip 5pt

Once the above simulation is set-up, we estimate the binwise cross-section 
$\chi^2(m_0,c_0)$ for each value of $m_0$ and $c_0$ and use this
distribution to calculate
\begin{equation}
\chi^2(m_0,c_0) ~~=~~ \sum_{i = 1}^{20}
\frac{ \left[ \sigma_i (m_0,c_0) - \sigma_i^{(C.V.)} \right]^2}
     {  \epsilon_i^2 }
\end{equation}
where
\begin{equation}
\epsilon_i =
\epsilon^i_1 ~\theta[ \sigma_i (m_0,c_0) - \sigma_i^{(C.V.)} ]
~~+~~ \epsilon^i_2 ~\theta[ \sigma_i^{(C.V.)} - \sigma_i (m_0,c_0) ]
\end{equation}
assuming that the experimental value in the $i$-th bin is given by
$ [\sigma_i^{(C.V.)}]^{+\epsilon^i_1}_{-\epsilon^i_2}$. In this,
it is assumed that $\epsilon^i_1$ and $\epsilon^i_2$ contain the
statistical and systematic errors added in quadrature. The 95\% C.L.
bound is then obtained by requiring $\chi^2(m_0,c_0) < 31.41$, which is
the expectation~\cite{PDG} from random fluctuations.

% ------------------------------------------------------------------
\begin{figure}[htb]
\begin{center}
\vspace*{2in}
      \relax\noindent\hskip -4.4in\relax{\includegraphics{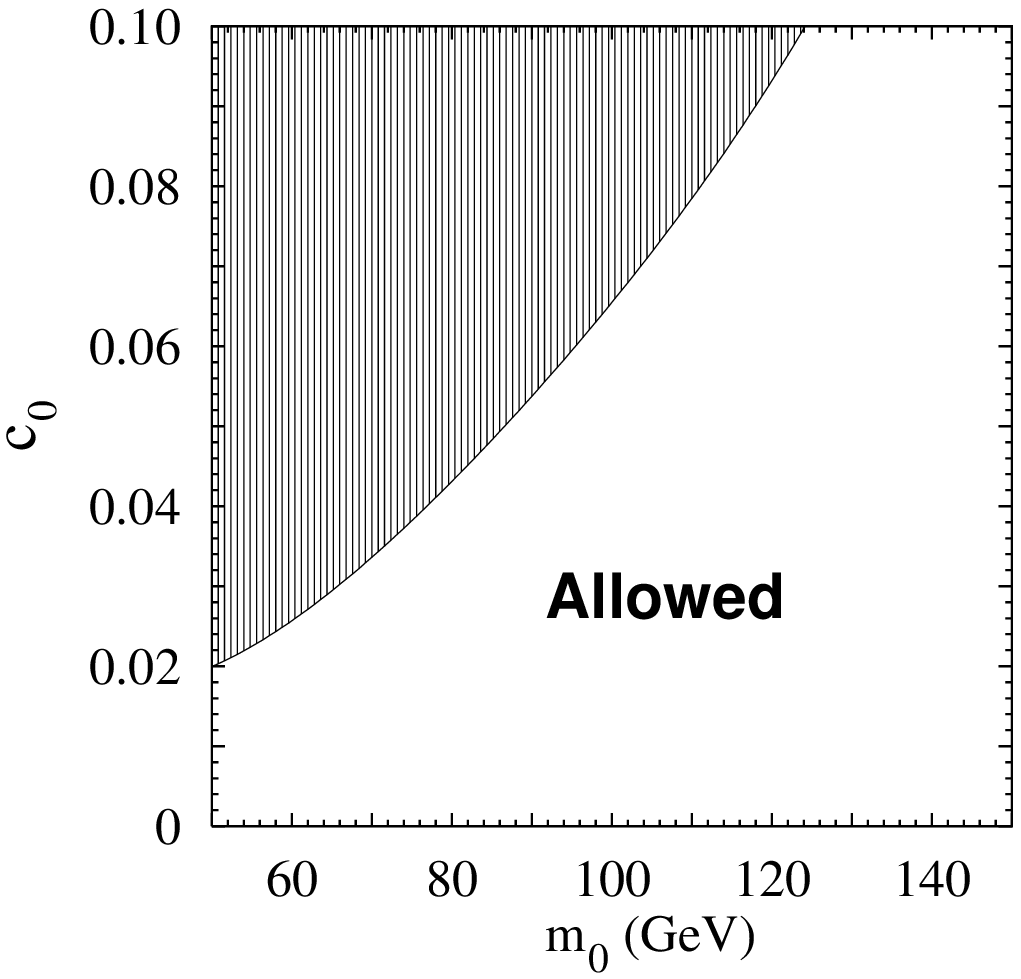}}
\end{center}
\end{figure}
\vspace*{1.3in}
\noindent {\bf Figure 2}.
{\footnotesize\it Illustrating the constraint on the parameter
space of the RS model arising from an analysis of ZEUS high-$Q^2$
data.  The shaded region is ruled out at the 95\% C.L. level. }
% ------------------------------------------------------------------
\vskip 5pt

In Figure 2, we show the 95\% C.L. constraints on the $m_0$--$c_0$ plane
using the above technique. Since the effective graviton coupling is 
quadratic in $c_0$ we expect the cross-section to rise as
$c_0$ increases --- this is reflected in the fact that Figure 2
shows upper bounds on $c_0$. On the other hand, the $m_0$ dependence
of the cross-section is very complicated, because of the summation over
the KK states. However, as the figure makes clear, there is a 
sharp drop in the cross-section as $m_0$ increases, so that the
ZEUS data become insensitive to the new physics beyond about 
$m_0 = 120$ GeV. This corresponds to $M_1 \simeq 460$ GeV, a value
which is still not accessible to the generation of colliders running
at present. 

As the above figure and discussion makes clear, HERA data as presented
by the ZEUS Collaboration provide
somewhat modest, but nevertheless interesting constraints on the
parameter space of the RS model of quantum gravity. Since gravitons
couple to the energy-momentum tensor of the matter fields, one may
expect considerable improvements in these results at machines running
at higher energies, such as the LHC, the proposed NLC and possible
muon colliders. In particular, it would be interesting to see if these
machines could actually find graviton resonances, which one expects
in the RS model~\cite{Davo-Hewt-Rizz}, 
but not in the ADD theory. We have performed a 
preliminary study of the RS model in the light of HERA data, and we
expect that the future will see many more detailed studies of this 
very interesting scenario.

\vskip 10pt
\noindent  
The authors have benefited greatly from discussions with Gour 
Bhattacharya, Dilip K. Ghosh,  Sudipta Mukherji, Gautam Sengupta,
K. Sridhar and Ajit M. Srivastava.  SS acknowledges financial support 
from the Council of Scientific and Industrial Research (Award No. 
9/92(185)/95-EMR-I), Government of India. 

%--------------------------------------------------------------------%

%%%%%%%%%%%%%%%%%%%%%%%%%%%%%%%%%%%%%%%%%%%%%%%%%%%%%%%%%%%%%%%%%%%%%%%%
\end{document}